\documentclass[a4paper, amsfonts, amssymb, amsmath, reprint, showkeys, nofootinbib,twoside]{revtex4-2}

\usepackage[english]{babel}
\usepackage[utf8]{inputenc}
\usepackage{soul}
\usepackage{titlesec}
\usepackage{tikz}
\usetikzlibrary{quantikz2}

\usepackage[colorinlistoftodos, color=green!40, prependcaption]{todonotes}
\usepackage{amsthm}
\usepackage{mathtools}
\usepackage{physics}
\usepackage{xcolor}
\usepackage{graphicx}
\usepackage[left=18mm,right=18mm,top=25mm,columnsep=15pt]{geometry} 
\usepackage{adjustbox}
\usepackage{placeins}
\usepackage[T1]{fontenc}
\usepackage{lipsum}
\usepackage{csquotes}
\usepackage{tikz}
\usepackage{svg}
\usepackage{float}
\usepackage{bm}
\usetikzlibrary{quantikz2}

\newtheorem{theorem}{Theorem}
\newtheorem{lemma}[theorem]{Lemma}       
\newtheorem{corollary}[theorem]{Corollary} 

\usepackage[ruled,lined]{algorithm2e}
\usepackage{algorithmic}

\usepackage{xspace}
\usepackage{hyperref}
\hypersetup{colorlinks,linkcolor=blue,anchorcolor=blue,citecolor=blue,urlcolor=blue}
\usepackage{bookmark}
\usepackage{csvsimple}
\usepackage{pgfplots}

\usetikzlibrary{shapes.geometric}
\usepackage{pgfplots}
\pgfplotsset{compat=newest}
\pgfdeclareplotmark{mystar}{
    \node[star,star point ratio=2.25,minimum size=6pt,
          inner sep=0pt,draw=violet,solid,fill=violet] {};
}

\newcommand{\prepare}{\mbox{P{\scriptsize REPARE}}}
\newcommand{\prep}{\mbox{P{\scriptsize REP}}}
\newcommand{\select}{\mbox{S\scriptsize ELECT}}

\begin{document}
\title{A shorter-width post-selected circuit for a truncated-Taylor Hamiltonian-evolution segment}

\author{Michelle Wynne Sze}
\email[]{michelle.sze@quantinuum.com}
\author{David Zsolt Manrique}
\email[]{david.zsolt.manrique@quantinuum.com}
\author{David Muñoz Ramo}
\email[]{david.munozramo@quantinuum.com}
\author{Nathan Fitzpatrick}
\email[]{nathan.fitzpatrick@quantinuum.com}
\affiliation{Quantinuum, 13-15 Hills Road, CB2 1NL, Cambridge, United Kingdom}
\date{\today} 

\begin{abstract}

We present an ancilla-efficient, post-selected circuit construction for a truncated-Taylor approximation to short-time Hamiltonian evolution. The construction combines a binary Taylor-order register with mid-circuit measurement, reset, and reuse of the Hamiltonian-index register. Conditioned on the prescribed all-zero outcomes, the circuit implements the truncated Taylor approximation while maintaining coherence among the Taylor-order branches while permitting early
termination after a failed post-selection. For Taylor cutoff \(K\), a Hamiltonian containing \(L\) terms, and an \(n\)-qubit system, the circuit uses $
\left\lceil\log {(K+1)}\right\rceil
+\left\lceil\log L\right\rceil+n $ qubits, and executes \(O(K)\) controlled Hamiltonian blocks in a complete
segment. We analyze its post-selection probability and expected per-segment
cost, and implement and evaluate the construction in
\texttt{Guppy}~\cite{Koch2024,Koch2025}. The present work focuses on the
low-width implementation of individual short-time propagation segments;
coherent composition into a deterministic long-time simulation requires an
additional amplification or compression mechanism and is left for future
work.

\end{abstract}

\maketitle
\label{sec:letter}

\section{Introduction}\label{sec:intro}
Hamiltonian dynamics simulations have been an active area of quantum algorithms research and development. Interesting quantum many-body systems come in large scales which can be challenging for a classical computer to simulate accurately. Hence the emergence of quantum algorithms to describe such systems more efficiently and less resource intensive. 

Central to the problem of Hamiltonian dynamics is tracking the evolution of a quantum system over time $t$. For a closed system, the evolution is governed by the operator $\exp(-iHt)$, where $H$ denotes the system Hamiltonian. Several quantum computational approaches exist for simulating this evolution. Here, we revisit the truncated Taylor series (TTS) method \cite{Berry_etal_2015} which utilizes the linear combination of unitaries (LCU)\cite{ChildsWiebe2012,berry_childs_cleve_kothari_somma_2017} algorithm.  
The time evolution operator is approximated to a desired accuracy $\epsilon$ by truncating its Taylor expansion at an order $K(\epsilon)$. The Hamiltonian $H$ is expressed as a sum of $L$ Pauli operators,
\begin{align}\label{eq:qubit_H}
    H = \sum_{\ell=0}^{L-1} \alpha_\ell H_\ell, \quad H_\ell = e^{i\theta_\ell} \bigotimes_{j=0}^{n-1} P_{\ell,j},
\end{align}
where $P_{\ell,j} \in \{I, X, Y, Z\}$, $\theta_\ell$ is some phase angle, and $n$ denotes the number of qubits in the system. 

Block-encoding methods \cite{Low2019hamiltonian,Chakraborty2019, ChildsWiebe2012,Gily_n_2019,Martyn2021GrandUnification}, such as the LCU, have recently gained much attention due to their better asymptotic scaling despite their overhead costs \cite{low2019hamiltoniansimulationinteractionpicture, vasconcelos2025methodsreducingancillaoverheadblock}. The LCU is a frequently used block-encoding primitive because of its near-optimal complexity dependence on the target precision $\epsilon$ \cite{ChildsWiebe2012,berry_childs_cleve_kothari_somma_2017}. This algorithm embeds a generally non-unitary operator in a larger unitary matrix using two oracles -- one which prepares the weights of the LCU and the other implements the unitary operators in sequence. 

The LCU framework thus emerges naturally in the truncated Taylor series method, as originally shown in Ref.~\cite{Berry_etal_2015}. Effectively, there are two nested LCUs being implemented -- one for the $H$ itself, and another corresponding to the truncated Taylor expansion whose constituent unitaries are the powers of $H$. An early work~\cite{Berry_etal_2015} established a simulation query complexity of $\mathcal{O}(KT)$ and a total gate cost of $\mathcal{O}(KT L(n + \log L))$, where $T = t\sum_\ell \alpha_\ell$, and $K =\frac{\log(T/\epsilon)}{\log \log(T/\epsilon)}$ . The corresponding circuit utilizes $K+1$ ancilla registers: $K$ registers encode $K$ copies of $\lbrace\alpha_\ell\rbrace$, while an additional register stores the Taylor expansion coefficients. The unary encoding of the Taylor coefficients strategically allows the use of $K$ singly-controlled Hamiltonian LCUs. Further techniques are later proposed to improve the simulation accuracy and query and gate complexities via an adaptive method\cite{Meister2022tailoringterm}, by appending an extra correction step\cite{Novo2017}, or by exploiting anticommuting terms\cite{Zhao_2021}.

The standard unary construction for the short-time propagator, $\exp(-iH\tau)$ uses one Taylor-order register together with $K$ Hamiltonian-index registers \cite{Berry_etal_2015}. 
Its width is $K + K\lceil \log{L} \rceil + n$ qubits. This width, rather than the asymptotic query complexity motivates the circuit construction studied in this work. We propose a circuit construction utilizing only two ancilla registers leading to fewer number of qubits while preserving the same asymptotic circuit depth and number of controlled-Hamiltonians. This is realized by applying a parallelization method and mid-circuit measurements. Mid-circuit measurements, or incoherent computations\cite{vasconcelos2025methodsreducingancillaoverheadblock}, is an intuitive and conservative approach to perform multiplication of block-encodings without using extra ancillas. Since every measurement indicates whether the right block of the larger unitary is being projected, the circuit is terminated and restarted instantly if a failed projection is detected. This reduces not only the number of qubits for implementing $H^k$ and possible memory error from the idling qubits, but also the resources and runtime of the quantum circuit. While measurement operations can take longer than gate operations in some quantum architectures, our method remains advantageous when the time saved by avoiding additional gates outweighs the cost of mid-circuit measurements. Reference~\cite{vasconcelos2025methodsreducingancillaoverheadblock}
 explicitly describes this repeated-measurement construction and contrasts it with coherent compression gadgets~\cite{low2019hamiltoniansimulationinteractionpicture}. Applied to truncated Taylor simulation, register reuse alone changes the width from $K + K\lceil \log L \rceil + n$ to $K + \lceil \log L \rceil + n$. It therefore accounts for the saving of $(K-1)\lceil \log L \rceil$ qubits.  

Different parallelization methods \cite{moore1998parallelquantumcomputationquantum,Zhang_2024,boyd2024lowoverheadparallelisationlcucommuting} can be exploited to tackle the expensive subroutine that uses multi-control operations in LCU. Here, we apply the technique discussed in Refs.~\cite{moore1998parallelquantumcomputationquantum,Zhang_2024}. In particular, the parallelization via binary encoding allows the preparation of Taylor coefficients using $\lceil \log (K+1) \rceil$ qubits without compromising the number of singly-controlled-$H$'s. Its incremental saving results in total width of $\lceil \log (K+1) \rceil + \lceil \log L \rceil + n$.

The two sources of reductions -- incoherent block-encoding multiplication and coherent binary power selection -- are apparently compatible. The Hamiltonian-index register is measured and reset while the Taylor-order register remains in a superposition over $k$. A measurement would decohere that superposition if its retained record distinguished the branches. Here it does not: every continuing branch is accepted on the same rank-one all-zero outcome. Post-selection may renormalize the branches, but it does not dephase them on the all-success trajectory. We prove this compatibility in Appendix and use this to obtain a hybrid coherent/post-selected truncated-Taylor circuit. The construction retains coherence in the binary Taylor register, recycles a single $\lceil \log L \rceil$-qubit Hamiltonian-index register, and aborts a shot when an intermediate all-zero test fails. A complete shot executes $\mathcal{O}(K)$ controlled Hamiltonian blocks. Early termination reduces the number of later blocks executed after a failure. We derive the joint success probability and the exact first-failure stopping-time expression, and evaluate compiled operation-count proxies for representative Ising-model Hamiltonians.

\section{Shorter width circuit}\label{sec:swTTS}

We present a circuit construction which has only two ancilla registers -- one for encoding the Taylor expansion coefficients, and the other one for the Hamiltonian coefficients. Mid-circuit measurements are also employed. 

Expanding $U_{\tau,K}=e^{-iH\tau}$ in a truncated Taylor series, 
\begin{align}
    U_{\tau,K} &= \sum_{k=0}^{K} \widetilde{\beta}_k \widetilde{H}^k, \text{ where} \\
    \widetilde{\beta}_k &= \frac{(\tau \left\|\alpha\right\|_{1})^k}{k!},  \text{  } \widetilde{H}= \frac{-i}{\left\|\alpha\right\|_{1}}\sum_{\ell}^{L}\alpha_\ell H_\ell,
\end{align}
and $\left\|\alpha\right\|_{1}=\sum_\ell |\alpha_\ell|$ is the $\ell_1$-norm of the coefficients $\{ \alpha_\ell \}$, we see that $U_{\tau,K}$  is a linear combination of moments of a rescaled Hamiltonian $\widetilde{H}^k$. We first set up the LCU for $\widetilde{H}^k$. 

\subsection{LCU for $\widetilde{H}^k$}
For $k=1$, the standard LCU construction is utilized employing the unitary oracles $\prepare(\alpha)$ and $\select(\widetilde{H})$ whose actions are given by
\begin{align}
\prepare(\alpha)&: |\bar{0}\rangle \mapsto \sum_{\ell=1}^{L} \sqrt{\frac{\alpha_\ell}{\left\|\alpha\right\|_1}} |\ell\rangle, \\
\select(\widetilde{H})&: |\ell\rangle |\psi\rangle \mapsto |\ell\rangle (-iH_\ell)|\psi\rangle.
\label{eq:selectHnorm}
\end{align}
$\prepare(\alpha)$ is implemented using $\lceil \log{L}\rceil$ qubits. 

For $k=2$, the circuit is constructed by composing two block-encodings. A straightforward implementation requires one extra ancilla qubit beyond the standard prepare register, and a set of $\lceil \log{L}\rceil$-open-controlled generalized Toffoli and X gates \cite{sünderhauf2023generalized,Dalzell_2025}. Post-selection is performed on the extra qubit at the end. This construction can be extended for $k>2$ with at most $k-1$ extra ancilla qubits and sets of $\lceil \log{L}\rceil$-open-controlled generalized Toffoli and X gates.
\begin{figure*}[htbp!]
    \centering
    \begin{tikzpicture}
    \node[scale=0.90] {
    \begin{quantikz}
            \lstick{$|\bar{0}\rangle_{\ell}$}&\gate[2]{W_{\widetilde{H}^k}} & \meterD{\bar{0}} \\
            \lstick{$|\psi\rangle$}&  &  \push{\rstick{$\widetilde{H}^k|\psi\rangle$}}
        \end{quantikz}
        =
        \begin{quantikz}
            \lstick{$|\bar{0}\rangle_\ell$}& \qwbundle{} & &\gate[2]{U_{\widetilde{H}}}\gategroup[2,steps=7,style={inner xsep=19pt,dashed,rounded corners},label style={label position=below, anchor=north,yshift=-0.2cm}]{$k$ $U_{\widetilde{H}}$+post-selection boxes} &  \meterD{\bar{0}} &    \gate[2]{U_{\widetilde{H}}} &  \meterD{\bar{0}} & \ \ldots\ &\gate[2]{U_{\widetilde{H}}}&  \meterD{\bar{0}}   \\
            \lstick{$|\psi\rangle$}& \qwbundle{} &  &  &  & & & \ \ldots\ &  & {\rstick{$\widetilde{H}^k|\psi\rangle$}}
    \end{quantikz}
    };
\end{tikzpicture}
\caption{\textbf{$\bm k$ LCU blocks of $\bm{\widetilde{H}}$ for $\bm{\widetilde{H}^k}$.} Post-selection is performed after every LCU $U_{\widetilde{H}}$. An unsuccessful post-selection aborts and restarts the circuit.
}
\label{fig:Hk_k_blocks}
\end{figure*}
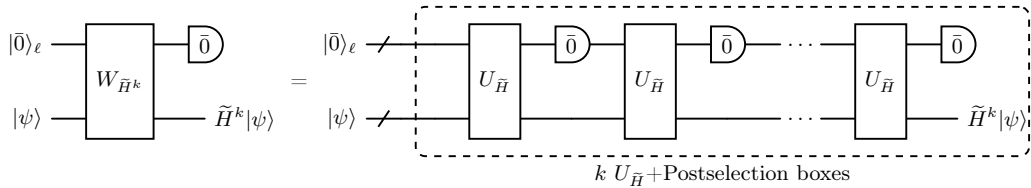

Alternatively, we use the identity\cite{sünderhauf2023generalized}
\begin{equation}
    \begin{quantikz}
    \lstick{$|0\rangle$} &  &\targ{0}  & \gate{X} &\meterD{0} \\
      &\qwbundle{} & \octrl{-1} & &  
    \end{quantikz}
        =
    \begin{quantikz}
        & & & \\
            & \qwbundle{} & \meterD{\bar{0}} & \qwbundle{}
    \end{quantikz}
\end{equation}
to bypass the use of $k-1$ extra ancilla qubits. A $0$-measurement on the top ancilla qubit indicates a measurement of all-$0$ at the bottom register which automatically resets to $|\bar{0}\rangle = |0\cdots0\rangle$ state. Thus, to simulate $\widetilde{H}^k |\psi \rangle$, we utilize an operator $W_{\widetilde{H}^k}$, which is a sequence of $k$ LCUs of $\widetilde{H}$, or $U_{\widetilde{H}}$, each followed by a measurement post-selection. This requires {\it at least} $k$ mid-circuit measurements. The action of $W_{\widetilde{H}^k}$ is given by
\begin{align}\label{eq:W_Hk}
   W_{\widetilde{H}^k}  |\bar{0} \rangle_\ell|\psi \rangle = |\bar{0} \rangle_\ell \widetilde{H}^k|\psi \rangle + |\bot\rangle_\ell |\phi \rangle,
\end{align}
where $|\bot\rangle_\ell |\phi \rangle$ is the state orthogonal to $|\bar{0} \rangle_\ell \widetilde{H}^k|\psi \rangle$. After post-selection, $\widetilde{H}^k |\psi \rangle$ is extracted. The proposed circuit for $W_{\widetilde{H}^k}$ is sketched in Fig.~\ref{fig:Hk_k_blocks}, and has a constant qubit count of $\lceil \log L\rceil + n$ for any $k$.

A successful post-selection must be obtained before proceeding from one LCU to the next; otherwise, the circuit is restarted. Let $p_i = \langle \psi_{i-1} | \widetilde{H}^\dagger \widetilde{H} |\psi_{i-1}\rangle $ be the success probability of implementing the $i$-th application of $U_{\widetilde{H}}$, where $|\psi_i\rangle$ is the normalized state after the $i$-th successful application.  Then, the success probability of implementing $W_{\widetilde{H}^k}$ is given by
\begin{align}
    p_k p_{k-1} ...p_2 p_1 = \langle\psi_0 |  (\widetilde{H}^k)^\dagger \widetilde{H}^k | \psi_0 \rangle,
    \label{eq:success_probability_Hk}
\end{align}
which is identical to the success probability obtained using deferred measurements with additional ancilla qubits.

\subsection{LCU for $U_{\tau,K}$}
We next construct an LCU representation of $U_{\tau,K}$. Using three quantum registers, we define the circuit
\begin{align}\label{eq:lcu_W2}
\widetilde{W} = (\widetilde{B}^\dagger \otimes \mathbb{I}) \select(\widetilde{V}) (\widetilde{B} \otimes \mathbb{I}),
\end{align}
where $\widetilde{B}$ is the state-preparation operator
\begin{align}
\widetilde{B} := \prepare(\widetilde{\beta}) : |\bar{0}\rangle_k \mapsto \sum_{k=0}^{K}
\sqrt{\frac{\widetilde{\beta}_k}{\|\widetilde{\beta}\|_1}}|k\rangle,
\end{align}
with $\|\widetilde{\beta}\|_1$ denoting the $\ell_1$-norm of the coefficients $\widetilde{\beta}_k$. The coefficients are encoded in binary within a control register $|k\rangle$ consisting of $\lceil \log K \rceil$ qubits.

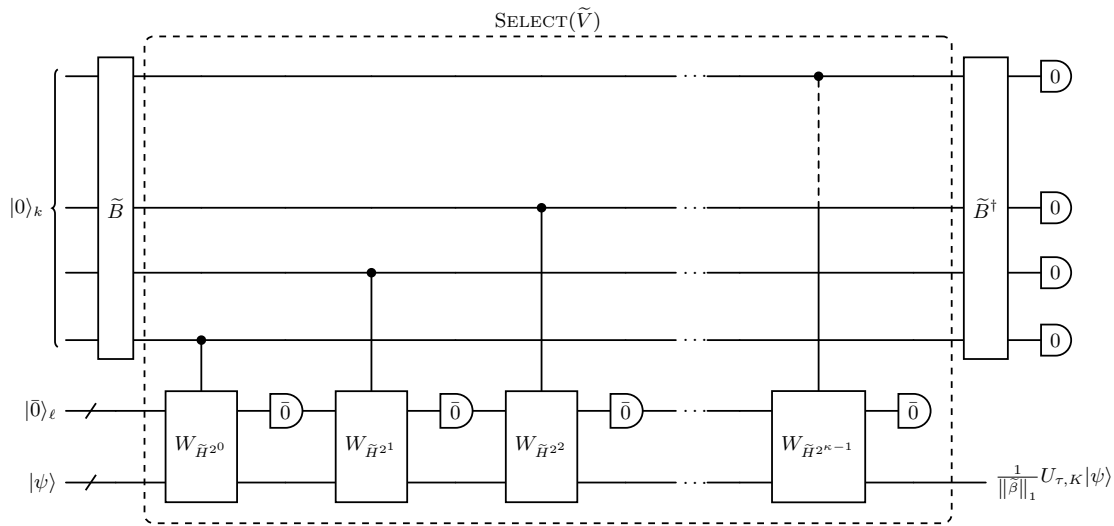
\begin{figure*}[htbp!]
\begin{tikzpicture}
\node[scale=0.85] {
\centering
\begin{quantikz} 
\lstick[4]{$|0\rangle_k$} &  \gate[4]{\widetilde{B}} &\gategroup[6,steps=10,style={dashed, rounded corners, inner sep=6pt}]{\select($\widetilde{V}$)}  & &  & &  & & \ \ldots & &\ctrl{0}\wire[d][1][style={dashed}]{q} & & \gate[4]{\widetilde{B}^\dagger} & \meterD{0} \\[1cm]
 & & & &  & & \ctrl{3}  & & \  \ldots & & \wire[d][1]{q} & & & \meterD{0}\\
& & & & \ctrl{2} & &   & &\  \ldots & &  \wire[d][1]{q}& & &\meterD{0}\\
& &\ctrl{1} & & & &   & &\  \ldots & & \wire[d][1]{q} & & &\meterD{0}\\
\lstick{$|\bar{0}\rangle_\ell$} & \qwbundle{} & \gate[2]{W_{\widetilde{H}^{2^0}}}& \meterD{\bar{0}} &\gate[2]{W_{\widetilde{H}^{2^1}}} & \meterD{\bar{0}} &\gate[2]{W_{\widetilde{H}^{2^2}}} & \meterD{\bar{0}} & \  \ldots &  &\gate[2]{W_{\widetilde{H}^{2^{\kappa-1}}}}& \meterD{\bar{0}}  \\
\lstick{$|\psi\rangle$} & \qwbundle{} &  & & & & &   & \  \ldots & & & & {\rstick{$\frac{1}{\left\|\widetilde{\beta}\right\|_1}U_{\tau,K}|\psi\rangle$}}
\end{quantikz}
};
\end{tikzpicture}
\caption{\textbf{Shorter width truncated Taylor series  expansion circuit}. $\widetilde{W}$ for $e^{-iH\tau}$ (up to order $K$). The $W_{\widetilde{H}^k}$ operator is defined in Fig.~\ref{fig:Hk_k_blocks}. $\kappa = \lceil \log (K+1) \rceil$ qubits are used to \prepare{}  the Taylor series coefficients and $K$ singly-controlled LCUs of $\widetilde{H}$ to implement the \select($\widetilde{V}$). }
\label{fig:singlyctrl_LCU}
\end{figure*}

The operator \(\select(\widetilde{V})\) implements a controlled application of the powers of $\widetilde{H}$. More specifically, it comprises a sequence of multi-controlled $W_{\widetilde{H}^k}$ operations and post-selection procedures, defined in Eq.~(\ref{eq:W_Hk}) and Fig.~\ref{fig:Hk_k_blocks}, acting on the $|\ell\rangle$ and $|\psi\rangle$ registers, with the control provided by the register $|k\rangle$. The corresponding success Kraus operator after measurement is 
\begin{align}
    M_0 = \sum_k |k \rangle \langle k| \otimes \widetilde{H}^k,
\end{align}
and its action is given by
\begin{equation}
\select(\widetilde{V}): |k\rangle |\ell\rangle |\psi\rangle
\mapsto
|k\rangle |\bar{0}\rangle \widetilde{H}^{k} |\psi\rangle.
\label{equation:selectVtilde}
\end{equation}
This is reminiscent of the linear combination of block encoded matrices articulated in Refs.~\cite{Childs_2017,Gily_n_2019}. Since $W_{\widetilde{H}^k}$ is a unitary, the multi-control scheme can be simplified to a singly-control one using the identity \cite{moore1998parallelquantumcomputationquantum,Zhang_2024,Rosenkranz2025quantumstate} 
\begin{widetext}
  \begin{align}
    \sum_{k=0}^{K} |k\rangle\langle k| \otimes U^k = \prod_{i=1}^{\kappa} \left[ |0\rangle\langle 0|_{i-1} \otimes I^{\otimes n} + |1\rangle\langle 1|_{i-1} \otimes U^{2^{i-1}}  \right],
\end{align}  
\end{widetext}
where $K = 2^\kappa -1$ and $U$ is any unitary operator. This leads to $\kappa$ singly-controlled $W_{\widetilde{H}^{2^i}}$, ($i=0,1,..., \kappa -1$), as in Fig.~\ref{fig:singlyctrl_LCU}, or $K$ ($=\sum_{i=0}^{\kappa-1} 2^i = 2^{\kappa}-1$) singly-controlled $U_{\widetilde{H}}$ and {\it at least} $K$ mid-circuit measurements. The $\widetilde{W}$-circuit in Fig.~\ref{fig:singlyctrl_LCU} utilizes the same number of controlled-$H$ as the circuit in Ref.~\cite{Berry_etal_2015}, but with fewer number of qubits. The application of both the repeated-measurement multiplication and binary controlled-power identity for truncated-Taylor LCU establishes that recyclable post-selection preserves the coherent superposition over Taylor order. This is proven in Lemma~\ref{lem:common-outcome} and the following Corollary~\ref{cor:recycled-powers} in the Appendix.

Applying $\widetilde{W}$ to the initial state yields
\begin{widetext}
 \begin{align}
    \widetilde{W}|\bar{0} \rangle_k |\bar{0} \rangle_\ell|\psi \rangle = \frac{1}{\left\|\widetilde{\beta}\right\|_1} |\bar{0} \rangle_k |\bar{0} \rangle_\ell U_{\tau, K} |\psi \rangle + \sqrt{1-\frac{1}{\left\|\widetilde{\beta}\right\|_1^2}}\left|\perp\rangle_{k,\ell,\psi}\right.,
\end{align}   
\end{widetext}
where $\left|\perp\rangle_{k,\ell,\psi}\right.$ represents the residual component orthogonal to the successful ancilla configuration  $|\bar{0}\rangle_k$, satisfying $\langle \bar{0}_k \left|\perp\rangle_{k,\ell,\psi}\right. = 0$. In Fig.~\ref{fig:singlyctrl_LCU}, $\widetilde{W}|\bar{0} \rangle_k |\bar{0} \rangle_\ell|\psi \rangle$ is the intermediate state obtained after applying $\widetilde{B}^\dagger$, following $K$ successful post-selections on $|\cdot\rangle_\ell$-register, and preceding post-selection on the $|\cdot\rangle_k$-register. 

\begin{figure}[htbp!]
    \centering
    \includesvg[width=3.2in]{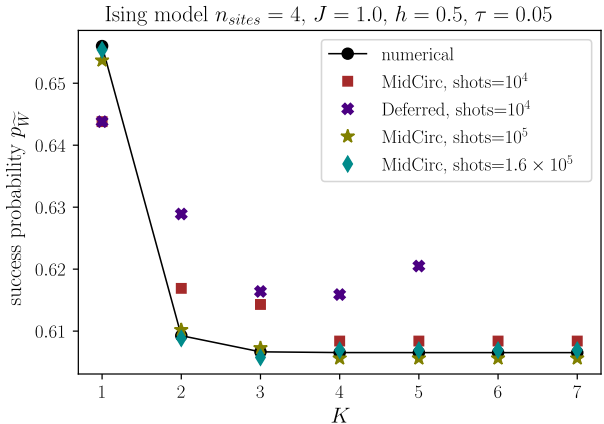}
\caption{\textbf{Success probability of simulating} $\bm{U_{\tau,K}}$. A 1D-four-lattice-site Ising model Hamiltonian, $H=J\sum_{\langle i,j\rangle}^n Z_i Z_j + h\sum_i^nX_i$, is used with interaction strength $J=1.0$ a.u., external magnetic field $h=0.5$ a.u., and $\tau=0.05$ a.u. All non-numerical data points are obtained using (statevector) simulations, in the absence of any error model, of the circuits defined by $\widetilde{W}$ (with mid-circuit measurements) and $W$ sketched in Fig.~\ref{fig:U_lcu_berry1} (with deferred measurements). For $\widetilde{W}$, a successful shot means having measured $K$ $|\bar{0}\rangle_\ell$ before $|\bar{0}\rangle_k$ at the end. The numerical data set is obtained by calculating the right-hand side of Eq.~(\ref{eq:success_probability_U}) via matrix multiplication. }
\label{fig:SuccessProbability_IsingPropagator_n=4}
\end{figure}

Finally, the success probability of projecting to the $|\bar{0}\rangle_k$ state is given by 
\begin{align}
    p_{\widetilde{W}} = \frac{\langle \psi | U_{\tau,K}^\dagger U_{\tau,K} |\psi \rangle }{\left\|\widetilde{\beta}\right\|_1^2},
\label{eq:success_probability_U}
\end{align}
which is equivalent to the result of Ref.~\cite{Berry_etal_2015}. We substantiate this by simulating the circuit for a 1D Ising model Hamiltonian with four lattice sites. A pseudocode for implementing $\widetilde{W}$ is provided in the Appendix, and actual script is available on \cite{github}. The programming language \texttt{Guppy}\cite{Koch2024,Koch2025} is utilized to enable the execution of circuits with mid-measurements and early-exits. In Fig.~\ref{fig:SuccessProbability_IsingPropagator_n=4}, we compare the success probabilities obtained by implementing $\widetilde{W}$, with varying number of shots, and $W$ in Fig.~\ref{fig:U_lcu_berry1}. Higher number of shots leads to faster convergence to the analytical results. Convergence with respect to $K$ is expected despite performing product of block encodings which would lead to exponentially decaying success probabilities $\langle\psi |  (\widetilde{H}^k)^\dagger \widetilde{H}^k | \psi \rangle$ due to its $1/\left\|\alpha\right\|_1^{2k}$ scaling. This scaling is canceled out in the corresponding $\widetilde{\beta}_k$ that gives weight to the controlled-$\widetilde{H}^k$. Moreover, with increasing $k$, $\widetilde{\beta}_k$ renders $\langle\psi |  (\widetilde{H}^k)^\dagger \widetilde{H}^k | \psi \rangle$ trivial because the corresponding weights become insignificant to the overall $U_{\tau,K}$.  

\subsection{Amplifying the success probability by optimizing the $\ell_1$-norm}
While the $\ell_1$-norm $\left\|\alpha\right\|_1$ contributes trivially to the block encoding of $U_{\tau,K}$, it is still crucial for implementing the first few controlled-Hamiltonian in $\widetilde{W}$. Given that circuits $\widetilde{W}$ and $W$ are equivalent, crucially via the circuit identity shown in Fig.~\ref{fig:lcu_Hk_normalized}, direct application of oblivious amplitude amplification (OAA) as described in Ref.~\cite{paetznick2014,Berry_etal_2015,Meister2022tailoringterm,Novo2017} to boost the success probability of the controlled-$W_{\widetilde{H}^k}$ may not be viable due to the mid-circuit post-selections. Some possible alternatives include leveraging equivalent block encoding circuits such as one in Fig.~\ref{fig:lcu_Hk_normalized}(middle) or one that uses addition compression gadget as proposed in Ref.~\cite{vasconcelos2025methodsreducingancillaoverheadblock, low2019hamiltoniansimulationinteractionpicture} -- all have deferred measurements but use fewer ancilla qubits than $W$, but still greater than $\widetilde{W}$.

\begin{table}[!htbp]
\centering
\begin{ruledtabular}
\begin{tabular}{l c c c c c c}
  \textbf{Molecule} & $\lVert\alpha\rVert_{1}$ & $\lVert\alpha\rVert_{1}'$ & $L$ & $L'$ & $p$ & $p'$ \\ 
\hline
BeH$_{\text{2}}$  & 30.219 & 21.912 & 666 & 1634 & 0.265 & 0.504 \\
CH$_{\text{4}}$  & 91.251 & 65.541 & 6892 & 7804 & 0.190 & 0.367 
 \\
 H$_{\text{2}}$O  & 118.420 & 88.021 & 1086 & 1898 & 0.401 & 0.725
 \\
 NH$_{\text{3}}$  & 100.122 & 72.832 & 3057 & 3889 & 0.307 & 0.580 
 \\
 F$_{\text{2}}$  & 296.81 & 234.35 & 2951 & 6063 & 0.436 & 0.699 
 \\
 HCl  & 605.98 & 514.54 & 5851 & 7975 & 0.564 & 0.782 
 \\
\end{tabular}
\end{ruledtabular}
\caption{\textbf{Improvement in the success probability} $\bm p$ of $U_{\widetilde{H}}$ due to $\ell_1$-norm $\lVert\alpha\rVert_1$ optimization of the Jordan-Wigner encoded Hamiltonian coefficients. Norms are in Hartree. $L$ is the number of terms in the Hamiltonian. The parameters $\lVert\alpha\rVert_1'$, $L'$, and $p'$ are the new $\lVert\alpha\rVert_{1}$, $L$, and $p$ after performing BLISS method. }
\label{tab:success_prob}
\end{table}

A route to improve the post-selection success probability is to reduce the $\ell_{1}$-norm of the Hamiltonian coefficients by exploiting symmetries and other invariant transformations~\cite{doi:10.1021/acs.jctc.3c00912, patel2025quantum}.  
To demonstrate how such symmetry based reductions can improve post-selection 
efficiency, we apply the block-invariant symmetry shift (BLISS) 
approach~\cite{doi:10.1021/acs.jctc.3c00912} to a few second-quantized molecular 
Hamiltonians, replacing $H$ with the energetically equivalent operator
\begin{align}
   H(\xi_0,\boldsymbol{\xi})
   =H-\Bigl(\xi_0+\sum_{i,j=0}^{N}\xi_{ij}\,a_{i}^{\dagger}a_{j}\Bigr)
        \bigl(\hat N_{e}-N_{e}\bigr),
\end{align}
where $a_{i}^{\dagger}$ ($a_{i}$) are the electronic creation (annihilation) operators, $\hat N_{e}=\sum_{k=0}^{N}a_{k}^{\dagger}a_{k}$ is the total electron number operator and $N_{e}$ is the electron number. The scalar $\xi_0$ and Hermitian matrix $\boldsymbol{\xi}$ are chosen such that, after Jordan--Wigner encoding, the coefficients $\{\alpha_\ell\}$ attain the minimum possible $\ell_{1}$-norm, hence increasing the LCU success probability $p$. To numerically quantify the gain, we combine BLISS with Jordan–Wigner encoding and optimize the Hamiltonians of the molecules listed in Table~\ref{tab:success_prob}. In every case, the optimized Hamiltonian coefficients reduce $\lVert\alpha\rVert_{1}$ and increase~$p$ relative to the unmodified Hamiltonian.

\section{Average runtime}\label{sec:ave_runtime}
When measurement times are comparable to gate-operation times, mid-circuit measurements can improve resource efficiency by allowing failed post-selection outcomes to be identified early. The circuit may then be aborted and restarted before completion, avoiding unnecessary gate executions and conserving computational resources. Before estimating the resources conserved when applying mid-circuit measurements in $\widetilde{W}$, we first consider the total runtimes of successful LCU circuit executions for $\widetilde{H}^k$ with and without mid-measurements.

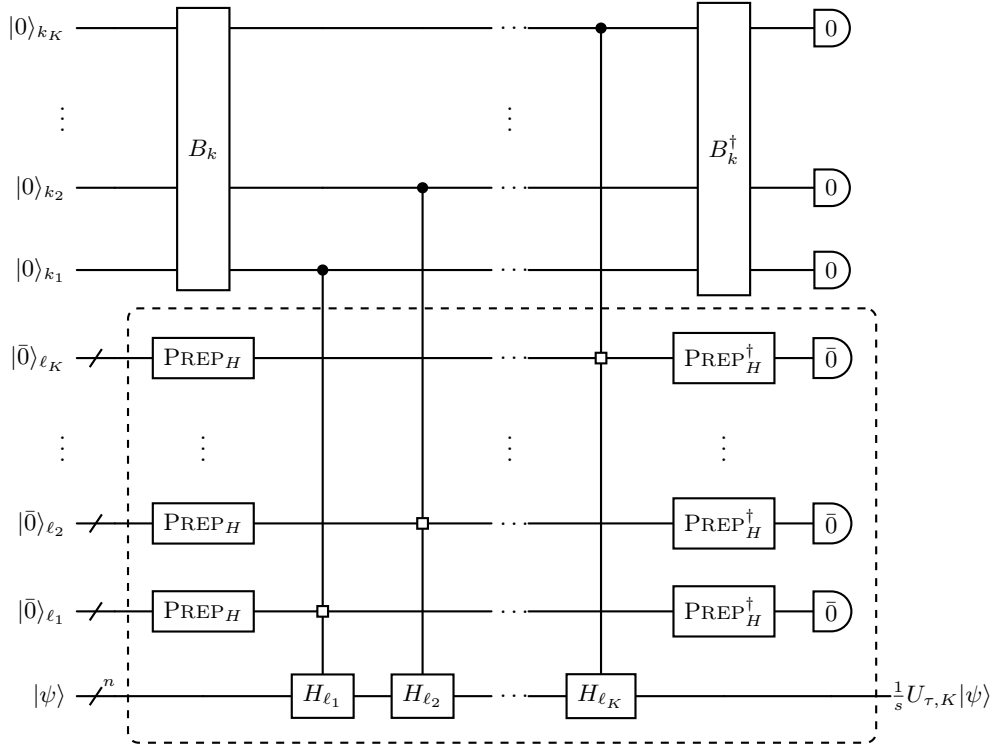
\begin{figure}[htbp!]
    \centering
    \begin{tikzpicture}
    \node[scale=1] {
    \begin{quantikz}[wire types={q,n,q,q,q,n,q,q,q}]%
    \lstick{$|0\rangle_{k_K}$}    &        &  \gate[4]{{B_k}}    &                         &                            & \ \ldots  & \ctrl{4}   & \gate[4]{{B_k^\dagger}} &\meterD{0}     \\  
    \lstick{$\vdots$}         &             &             &            &                            & \vdots    &            &   \\
    \lstick{$|0\rangle_{k_2}$}    &      &       &                         &    \ctrl{4}                & \ \ldots  &            &   &\meterD{0}\\
    \lstick{$|0\rangle_{k_1}$}    &    &         & \ctrl{4}                &                            & \ \ldots  &            &  &\meterD{0} \\
    \lstick{$|\bar{0}\rangle_{\ell_K}$} & \qwbundle{} &\gate{\prep_H} \gategroup[5,steps=7,style={dashed, rounded corners, inner sep=6pt}]{}                   &    &                            & \ \ldots  &|[operator]|& \gate{\prep^\dagger_H}  &\meterD{\bar{0}} \\  
    \lstick{$\vdots$}         &  & \vdots   &        &                         &                      \vdots      &     &     \vdots       &  &   \\
    \lstick{$|\bar{0}\rangle_{\ell_2}$} & \qwbundle{} & \gate{\prep_H}                 &       &  |[operator]|              & \ \ldots  &            & \gate{\prep^\dagger_H} & \meterD{\bar{0}} \\  
    \lstick{$|\bar{0}\rangle_{\ell_1}$} & \qwbundle{} & \gate{\prep_H} & |[operator]|             &                            & \ \ldots  &            &  \gate{\prep^\dagger_H} & \meterD{\bar{0}} \\
    \lstick{$|\psi\rangle$}   & \qwbundle{n} &  &\gate{H_{\ell_1}}\wire[u]{q}& \gate{{H_{\ell_2}}}\wire[u][2]{q}& \ \ldots  &\gate{H_{\ell_K}}\wire[u][4]{q}& & &{\frac{1}{s}{U_{\tau,K}|\psi\rangle}}
    \end{quantikz}
    };
\end{tikzpicture}
\caption{\textbf{LCU circuit $\bm W$ for $\bm{U_{\tau,K}}$}, where the Taylor expansion coefficients are encoded in unary as described in Ref.~\cite{Berry_etal_2015}. $B_k$ and $B_k^\dagger$ operators prepare the $K+1$ rescaled Taylor expansion coefficients using $K$ qubits. $\prep_H$ and $\prep^\dagger_H$ prepare the Hamiltonian coefficients $\alpha_\ell$. The $\select$ operator consists of $K$ singly-controlled-$H$'s. A successful post-selection on $K + K\lceil \log L \rceil$ qubits at the end of the circuit leads to $\frac{1}{s}U_{\tau,K}|\psi\rangle$, where $s=\lVert\widetilde{\beta}\rVert_{1}$ . The boxed subcircuit (without the single controls on $H$) performs the product of $K$ block encodings of $H$, where the end result upon successful post-selection on $K$ prepare registers leads to $\widetilde{H}^K|\psi\rangle = \frac{(-iH)^K}{\lVert\alpha\rVert_{1}^{2K}}|\psi\rangle$.}
\label{fig:U_lcu_berry1}
\end{figure}

\begin{figure*}[htbp!]
    \centering
    \begin{tikzpicture}
    \node[scale=0.7] {
     \begin{quantikz}
            \lstick{$|\bar{0}\rangle_{\ell}$}&\gate[2]{W_{\widetilde{H}^k}} & \meterD{\bar{0}} \\
            \lstick{$|\psi\rangle$}&  &  \push{\rstick{$\widetilde{H}^k|\psi\rangle$}}
        \end{quantikz}
        =
    \begin{quantikz}[wire types={q,n,q,q,q}]%
    \lstick{$|0\rangle_{k-1}$}      &             &                          &           &                         & \ \ldots  & \targ{0}  &   \gate{X}               &\meterD{0}\\  
    \lstick{$\vdots$}               &             &                          &           &                         & \vdots    &           &   \vdots               & \vdots \\
    \lstick{$|0\rangle_1$}          &             &                          & \targ{0}  & \gate{X}                & \ \ldots  &           &                          &\meterD{0}  \\
    \lstick{$|\bar{0}\rangle_\ell$} & \qwbundle{} &\gate[2]{U_{\widetilde{H}}}\gategroup[2,steps=6,style={inner sep=2pt,dashed,rounded corners},label style={label position=below, anchor=north,yshift=-0.2cm}]{$k$ $U_{\widetilde{H}}$}  & \octrl{-1}& \gate[2]{U_{\widetilde{H}}}& \ \ldots & \octrl{-3}& \gate[2]{U_{\widetilde{H}}}&\meterD{0\cdots0} \\
    \lstick{$|\psi\rangle$}         & \qwbundle{} &                          &           &                          &  \ \ldots&           & & \push{\rstick{${\widetilde{H}}^k|\psi\rangle$}} 
    \end{quantikz} 
    =
    \begin{quantikz}[wire types={q,n,q,q,q}]%
    \lstick{$|\bar{0}\rangle_{\ell_k}$} & \qwbundle{} &\gate{\prep_H} {}                   &    &                            & \ \ldots  &|[operator]|& \gate{\prep^\dagger_H}  &\meterD{\bar{0}} \\  
    \lstick{$\vdots$}         &  & \vdots   &        &                         &                      \vdots      &     &     \vdots       &  &   \\
    \lstick{$|\bar{0}\rangle_{\ell_2}$} & \qwbundle{} & \gate{\prep_H}                 &       &  |[operator]|              & \ \ldots  &            & \gate{\prep^\dagger_H} & \meterD{\bar{0}} \\  
    \lstick{$|\bar{0}\rangle_{\ell_1}$} & \qwbundle{} & \gate{\prep_H} & |[operator]|             &                            & \ \ldots  &            &  \gate{\prep^\dagger_H} & \meterD{\bar{0}} \\
    \lstick{$|\psi\rangle$}   & \qwbundle{n} &  &\gate{H_{\ell_1}}\wire[u]{q}& \gate{H_{\ell_2}}\wire[u][2]{q}& \ \ldots  &\gate{H_{\ell_k}}\wire[u][4]{q}&   \push{\rstick{$\widetilde{H}^k|\psi\rangle$}}
    \end{quantikz}
    };
\end{tikzpicture}
\caption{\textbf{Circuit for product of $\bm k$ block encoded $\bm{\widetilde{H}}$.} The $W_{\widetilde{H}^k}$ circuits in Fig.~\ref{fig:singlyctrl_LCU} can be replaced by the circuit on the left, transform the binary encoding of the Taylor expansion coefficients into unary, and obtain the circuit in Fig.~\ref{fig:U_lcu_berry1}.}
\label{fig:lcu_Hk_normalized}
\end{figure*}
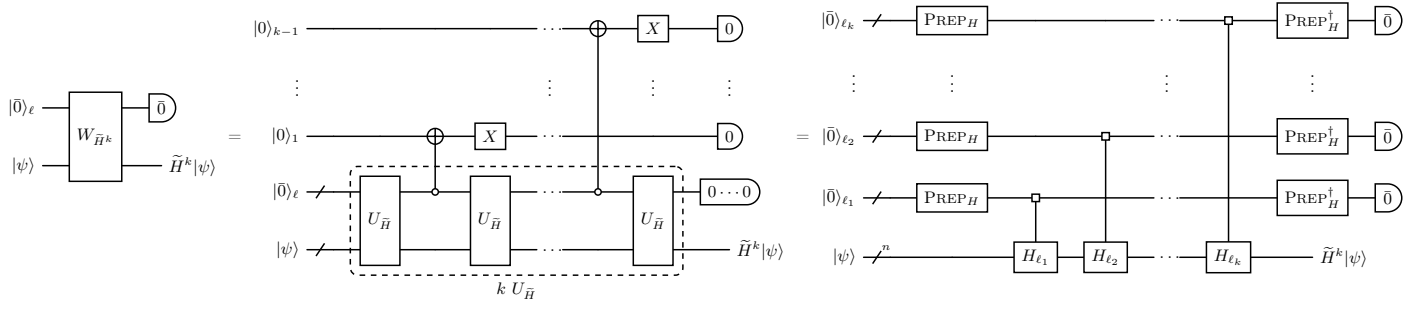

For the implementation of $\widetilde{H}^k$ with deferred measurements as captured in the {\it boxed} subcircuit in Fig.~\ref{fig:U_lcu_berry1}, the total runtime of a successful experiment is defined as the ratio of the average runtime of a single shot experiment to the success probability of projecting to $\widetilde{H}^k|\psi\rangle$. The success probability $\langle \psi | (\widetilde{H}^\dagger)^k \widetilde{H}^k|\psi \rangle$ has a lower bound $(\lambda_0/\left\|\alpha\right\|
_1)^{2k}$, where $\lambda_0\neq0$ is the smallest magnitude eigenvalue of $H$.  If $d$ is the average runtime of a single $U_H$, then the total runtime of $k$ $U_H$ with successful post-selections towards the end is given by $kd/\langle \psi | (\widetilde{H}^\dagger)^k \widetilde{H}^k|\psi \rangle$ which has an upper bound of $ kd(\left\|\alpha\right\|
_1/\lambda_0)^{2k}$. 

On the other hand, with mid-circuit measurements as in Fig.~\ref{fig:Hk_k_blocks}, the average runtime of a single shot experiment is the weighted sum of all possible runtimes of early terminated events and a complete one, that is
\begin{widetext}
\begin{align}
    \left[ \sum_{j=1}^{k-1} (1-p_j) \prod_{i=1}^{j-1}p_i jd\right] + p_1p_2...p_{k-1} kd = (1-p_1) d  + p_1 (1-p_2) 2d + p_1 p_2 (1-p_3) 3d +... + p_1 p_2...p_{k-1} kd  \nonumber
\end{align}
\end{widetext}
where we have used $p_i = \langle \psi_{i-1}| \widetilde{H}^\dagger \widetilde{H}|\psi_{i-1}\rangle$, with $|\psi_i\rangle$ a normalized state, and the terms in bracket of the left-hand side account for all possible early termination scenarios. Then, the total runtime of a successful experiment is given by
\begin{align}
     \frac{d(1 + p_1 + p_1 p_2 + ... + p_1p_2...p_{k-1})}{p_1 p_2...p_k}  \leq \frac{kd}{\langle \psi | (\widetilde{H}^\dagger)^k \widetilde{H}^k|\psi \rangle}.
     \label{eq:runtime_Hk}
\end{align}
Thus, the mid-circuit measurements provide greater resource savings in cases where the post-selection probability $\langle \psi | (\widetilde{H}^\dagger)^k \widetilde{H}^k|\psi \rangle$ is low. The equality in Eq.~(\ref{eq:runtime_Hk}) holds if $\widetilde{H}$ is a unitary or $p_1 = ... = p_{k-1}=1$. Otherwise, the inequality is more likely to hold.
\begin{figure}[htbp!]
    \centering
    \includesvg[width=3.2in]{figures/TwoQGateCountPropagatorLCU.svg}
\caption{\textbf{Average two-qubit gate, or weighted operation counts in $\bm W$ ($\circ$) and $\bm{\widetilde{W}}$ ($\star$) circuits.} The LCU circuits are compiled and run using the multiplexor synthesis technique demonstrated in Ref.~\cite{sze2025hamiltoniandynamicssimulationusing}. The $\star$ data points also include the number of mid-circuit measurements. The 1D-Ising model Hamiltonian, $H=J\sum_{\langle i,j\rangle}^n Z_i Z_j + h\sum_i^nX_i$, with varying lattice sites $n_{sites}$ are used. $W$ circuit utilizes linear amount of resources with increasing $K$, while it is piecewise linear in $\widetilde{W}$. $K$ values that share the same $\lceil \log(K+1) \rceil$ values (e.g. $K=2,3$ or $K=4,5,6,7$) require the same number of $K$ controlled-$U_{\widetilde{H}}$ operations. Consequently, for values of $K$ satisfying $\log(K+1) < \lceil \log(K+1) \rceil$, employing the $\widetilde{W}$ circuit yields no reduction in resource cost and can, in fact, be more expensive.   }
\label{fig:TwoQGateCount}
\end{figure}

\begin{figure*}[htbp!]
    \centering
    \includesvg[width=4.5in]{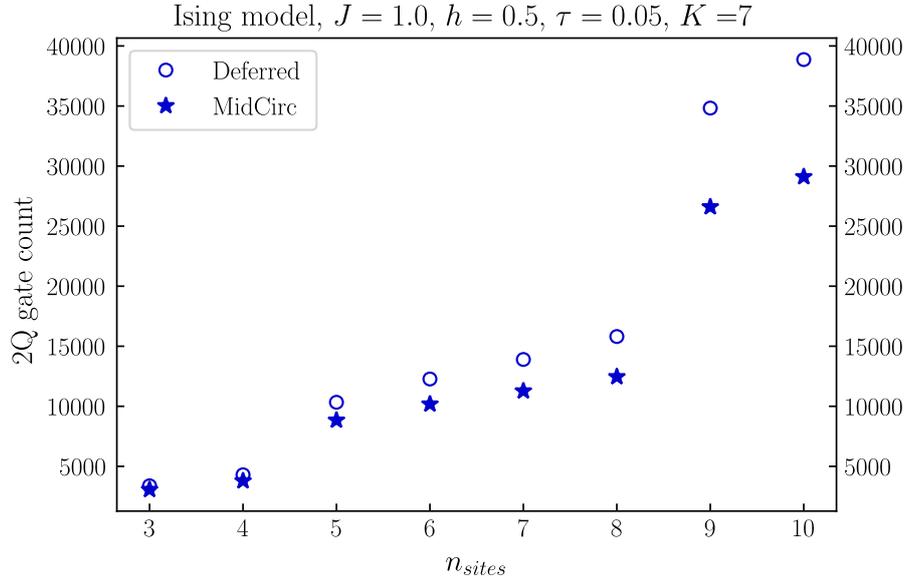}
\caption{\textbf{Two-qubit gates, or weighted operation counts, utilized in the $\bm W$ ($\circ$) and $\bm{\widetilde{W}}$ ($\star$) circuits as functions of system size $\bm {n_{sites}}$ for high-order $\bm K$.} The advantage of employing mid-circuit measurements becomes more pronounced as the system size becomes larger. The 1D-Ising model Hamiltonian, $H=J\sum_{\langle i,j\rangle}^n Z_i Z_j + h\sum_i^nX_i$, is used.}
\label{fig:TwoQGateCountK7}
\end{figure*}

As for the sequence of controlled-$U_{\widetilde{H}}$ in Fig.~{\ref{fig:singlyctrl_LCU}}, the effect of a mid-circuit post-selection  depends on the associated weight $\widetilde{\beta}_k$ on the controls. If $\widetilde{d}$ is the average runtime of a single controlled-$U_{H}$ together with its associated measurement, then the total runtime of $K$ controlled-$U_{H}$ with successful deferred post-selections is $K\widetilde{d}/p_{\widetilde{W}}$. With mid-circuit measurements, the first order or largest reduction in the total runtime of a successful implementation of $\widetilde{W}$ comes from the first failed (controlled-$U_{\widetilde{H}}$) post-selection, which has a probability of
\begin{align}
    f_1 = \sum_{k=0}^{\frac{K-1}{2}} \frac{\widetilde{\beta}_{2k+1}}{\left\|\widetilde{\beta}\right\|_1}(1-p_1) = \frac{1-p_1}{\left\|\widetilde{\beta}\right\|_1}\sum_{k=0}^{\frac{K-1}{2}} \frac{(\tau \left\|\alpha\right\|_{1})^{2k+1}}{(2k+1)!}; \nonumber
\end{align}
the first controlled block can fail only in Taylor branches with odd $k$. Let $f_j$ be the probability of first failed intermediate post-selection detected at block $j$. Then the average runtime $RT_{ave}$ of $\select (\widetilde{V})$ is given by 
\begin{align}
    RT_{ave} &= \frac{\widetilde{d}}{p_{\widetilde{W}}} \left[ K - \sum_{j=1}^{K-1} (K -j)f_j \right] \\
    & \leq \frac{K \widetilde{d}}{p_{\widetilde{W}}} \left[ 1 - \left(1-\frac{1}{K} \right)f_1 \right] \nonumber\\
    &\leq \frac{K\widetilde{d}}{p_{\widetilde{W}}}\left[ 1 - \left(1 - \frac{1}{K}\right)\frac{\tau \left\|\alpha\right\|_{1}}{\left\|\widetilde{\beta}\right\|_1}(1-p_1) \right].
\end{align}
The factor $1 - 1/K$ is required because a failure in the first block can only be detected after that block has run. Such a failure saves $K-1$ blocks, not all $K$ blocks. Since failed post-selection events are detected earlier, reduced expected runtimes are associated with smaller success probability $\langle \psi | \widetilde{H}^\dagger \widetilde{H}|\psi \rangle$. In Fig.~\ref{fig:TwoQGateCount}, we quantify the average runtime in terms of two-qubit gate count depth plus measurements for an Ising model Hamiltonian with different number of lattice sites. These counts should not be interpreted as hardware-independent wall-clock runtimes: their conversion to runtime depends on the relative durations of two-qubit gates, measurements, resets, classical feed-forward, and shot reinitialization.
The resource utilized in executing $\widetilde{W}$ is stepwise constant as a function of $K$ between successive powers of two, and the asymptotic gate complexity remains $\mathcal{O}(K)$. (Observe the $n_{sites}=3,4$ in the plot).  The scaling in two-qubit gate operations from compiling the circuit $W$ is linear with increasing $K$. Comparing the two sets of data, we observe that the savings from performing mid-circuit measurements becomes more apparent for higher $K$. (See also Fig.~\ref{fig:TwoQGateCountK7}.) Also, if $K$ satisfies $\log(K+1) < \lceil \log(K+1) \rceil$, then it may be more advantageous to consider the corresponding $\lceil \log(K+1) \rceil$ as this will provide better precision while utilizing the same number of controlled-$U_{\widetilde{H}}$. The data in Figs.~\ref{fig:TwoQGateCount} and~\ref{fig:TwoQGateCountK7} show that, for the Ising instances, synthesis method, and operation-cost convention considered here, early termination reduces the average executed operation count. The absolute saving can increase with Taylor order or system size because more operations remain to be skipped after an early failure. This trend is empirical rather than universal: the relative advantage is determined by the complete first-failure distribution $\lbrace f_j \rbrace$ and by the hardware-dependent costs of gates, measurements, resets, and feed-forward.

\section{Discussion and Conclusion}\label{sec:discussion_conclusion}
We have presented an alternative post-selected circuit for implementing a truncated-Taylor approximation to a short-time evolution operator. The
construction uses $\lceil\log(K+1)\rceil+\lceil\log L\rceil$ ancilla qubits, scaling
logarithmically with both the Taylor cutoff $K$ and the number $L$ of
Hamiltonian terms, while retaining the $\mathcal{O}(K)$ controlled-Hamiltonian-block
count of the original construction for a complete segment. Because oblivious
amplitude amplification is not directly applicable in the presence of the
intermediate measurements, the present construction is a probabilistic
short-time propagation protocol rather than a direct replacement for the
amplified multi-segment algorithm of Ref.~\cite{Berry_etal_2015}. Within this
post-selected setting, the BLISS-based reduction of the Hamiltonian coefficient
$\ell_1$-norm improves the per-segment acceptance probability. Moreover, when the
costs of measurement, reset, and classical feed-forward are sufficiently
small, terminating an unsuccessful execution at its first failed
post-selection reduces the expected number of operations performed per
attempted segment.

A major limitation of the original TTS algorithm is its qubit requirement, which remains one of the most costly resources in fault-tolerant quantum computing. Consequently, our width-reduction approach alleviates the need for substantial fault-tolerant infrastructure by lowering the overall qubit footprint.

While the incorporation of mid-circuit measurements can reduce circuit runtime through early termination and restart strategies, the proposed method does not fully eliminate the severe decay in success probability that arises at longer simulation times. This limitation stems from the fact that OAA is not immediately applicable within our framework. Moreover, employing OAA introduces significant additional circuit depth, leading not only to longer execution times but also to increased exposure to noise and logical errors.

Given the limited availability of reliable logical qubits and the ongoing development of quantum error-correction technologies, we do not anticipate the practical implementation of large-scale TTS algorithms with full OAA in the near term. Under these conditions, reducing circuit width becomes a more critical objective than maintaining a high success probability, particularly when qubit availability is the dominant resource constraint and the reduction in success probability can be compensated through inexpensive restart procedures enabled by mid-circuit measurements.

Therefore, we envision the post-selected protocol as being particularly well suited to the near-term to early fault-tolerant regimes, where short- to intermediate-time dynamics are of primary interest and the repetition overhead associated with probabilistic execution remains smaller than the resource savings obtained through circuit width reduction. This short-time propagation protocol may be useful for studying quantum Zeno effect, decoherence, early-times correlations with environment, or for short-time quantities like derivatives of observables, calculated as finite differences of the expected values of operators in short time evolutions. 

Compared with other block-encoding-based Hamiltonian simulation algorithms, such as those based on the quantum singular value transformation (QSVT), our modified TTS approach remains less attractive in the long term despite its reduced qubit requirements. QSVT achieves optimal asymptotic query complexity and circuit depth, making it the preferred framework for large-scale fault-tolerant quantum simulation. The primary obstacle to near-term implementation of QSVT is the need for long, coherent quantum signal processing (QSP) sequences, which remain challenging for current noisy hardware. However, QSVT is fundamentally designed for fault-tolerant execution and is expected to become the dominant approach as quantum hardware matures. In this context, the postselected short-time propagation protocol may serve as a practical intermediate solution, offering meaningful reductions in qubit requirements during the transition toward fully fault-tolerant quantum computation.

{\it Acknowledgements.--} We thank Marcello Benedetti, Ben Criger, Yuta Kikuchi and Gregory Boyd for helpful discussions, and the team behind \texttt{Guppy} programming language for the generous technical support.


\bibliography{references.bib}

\onecolumngrid
\section{Appendix} \label{sec:appendix_figures}

Measuring and restting the $\ell$-register must not reveal which Taylor-order branch $k$ is present.
\begin{lemma}[Common-outcome post-selection]
\label{lem:common-outcome}
Let \(C\) be a control register with computational basis
\(\{\ket{k}\}\), let \(A\) be an ancilla initialized in
\(\ket{0}_A\), and let \(S\) be the system.  Consider the controlled
unitary
\[
V=\sum_k \ket{k}\!\bra{k}_C\otimes U_k,
\]
where each \(U_k\) acts on \(A\otimes S\).  Define the successful
branch operator
\[
A_k=(\bra{0}_A\otimes I_S)\,U_k\,
    (\ket{0}_A\otimes I_S).
\]
If \(A\) is measured and the same outcome \(0\) is accepted for every
control branch, then the corresponding trace-nonincreasing map on
\(C\otimes S\) has the single Kraus operator
\[
M_0=\sum_k\ket{k}\!\bra{k}_C\otimes A_k.
\]
In particular, the accepted measurement record introduces no
which-\(k\) dephasing.
\end{lemma}

\begin{proof}
The Kraus operator associated with the accepted outcome is
\[
\begin{aligned}
M_0
&=(\bra{0}_A\otimes I_{CS})V
  (\ket{0}_A\otimes I_{CS})  \\
&=\sum_k\ket{k}\!\bra{k}_C\otimes A_k.
\end{aligned}
\]
Consequently, an off-diagonal block transforms as
\[
\ket{k}\!\bra{k'}\otimes X
\longmapsto
\ket{k}\!\bra{k'}\otimes A_k X A_{k'}^\dagger,
\]
rather than being projected to zero.  Thus the common accepted outcome
does not reveal the value of \(k\).
\end{proof}
In branch $k$, the successful component of the ancilla–system state is $\ket{0}_A A_k\ket{\psi_k}$. Post-selection retains the same pointer state $\ket{0}_A$ in every branch, so the measurement operation carries no record of $k$. The amplitudes of different branches may be changed by the operators $A_k$, but they remain part of one coherent conditional state.

\begin{corollary}[Recycled block encodings under binary control]
\label{cor:recycled-powers}
Suppose \(U_{\widetilde H}\) block encodes \(\widetilde H\), and in
round \(r\) it is applied when \(b_r(k)=1\) and omitted when
\(b_r(k)=0\).  If the block-encoding ancilla is post-selected on the
same all-zero outcome and reused after every round, then the Kraus
operator of the all-success trajectory is
\[
M_{\mathrm{succ}}
=
\sum_k\ket{k}\!\bra{k}
\otimes
\widetilde H^{\,\sum_r b_r(k)}.
\]
For the binary schedule in which bit \(k_i\) controls \(2^i\)
applications,
\[
\sum_r b_r(k)=\sum_i2^ik_i=k,
\]
and hence
\[
M_{\mathrm{succ}}
=
\sum_k\ket{k}\!\bra{k}\otimes\widetilde H^k.
\]
\end{corollary}

\clearpage

\begin{algorithm}
\caption{Truncated Taylor series for the time evolution operator with mid-circuit measurements}
\begin{algorithmic}[1]
    \STATE Specify $\kappa$, $N=\text{no. of shots}$
    \STATE Initialize the qubits of $k$-register to $|\bar{0}\rangle_k$ and the system to $|\psi\rangle$
    \STATE $i=N$
\WHILE{$i>0$}
    \STATE $i \gets i - 1$
    \STATE Perform $\widetilde{B}$ on $|\bar{0}\rangle_k$
    \FOR{$k=0$ to $\kappa-1$}
        \FOR{$k_j=0$ to $2^k-1$}
            \STATE Initialize the qubits of $\ell$-register to $|\bar{0}\rangle_\ell$
            \STATE Perform controlled-$U_{\widetilde{H}}$ \COMMENT{where control is at the $k$-th qubit of the $k$-register, and $U_{\widetilde{H}}$ is applied to $|\bar{0}\rangle_\ell |\cdot\rangle_\psi$}
            \STATE Measure $|\cdot\rangle_\ell$
            \IF{$|\cdot\rangle_\ell \neq |\bar{0}\rangle_\ell$}
                \STATE \textbf{break} \COMMENT{Exit loops early and restart}
            \ENDIF
        \ENDFOR
    \ENDFOR
    \STATE Perform $\widetilde{B}^\dagger$ on $|\cdot\rangle_k$
    \STATE Measure $|\cdot\rangle_k$
    \IF{$|\cdot\rangle_k = |\bar{0}\rangle_k$}
        \STATE Measure the new state $|\psi'\rangle$ \COMMENT{Successful projection. Keep this state}
    \ENDIF
\ENDWHILE
\end{algorithmic}
\end{algorithm}

\end{document}